# Understanding users' negative emotions and continuous usage intention in short video platforms

Xusen Cheng [a], Xiaowei Su [a], Bo Yang [a,*], Alex Zarifis [b], Jian Mou [c]

[a] *School of Information, Renmin University of China, China*
[b] *Cambridge Centre for Alternative Finance, University of Cambridge, UK; School of Business, University of Nicosia, Cyprus*
[c] *School of Business, Pusan National University, Republic of Korea*

| ARTICLE INFO | ABSTRACT |
| --- | --- |
| *Keywords:*<br>Short video<br>Flow theory<br>Illusion of control theory<br>Mixed research<br>Negative emotion<br>Continuous usage intention | While short videos bring a lot of information and happiness to users, they also occupy users' time and short videos gradually change people's living habits. This paper studies the negative effects and negative emotions of users caused by using short video platforms, as well as the users' intention to continue using the short video platform when they have negative emotions. Therefore, this study uses flow theory and illusion of control theory to construct a research hypothesis model and preliminarily confirms six influencing factors, and uses sequential mixed research method to conduct quantitative and qualitative research. The results show that users' use of short video platforms will have negative emotions and negative emotions will affect users' intention to continue to use short video platforms. This study expands the breadth and depth of research on short videos and enriches the research of negative emotions on the intention to continue using human–computer interaction software. Additionally, illusion of control theory is introduced into the field of human–computer interaction for the first time, which enriches the application scenarios of control illusion theory. |

## 1. Introduction

The Internet has changed the way people interact and reshaped traditional communication media such as film, television, music and the telephone (Statista, 2022). The emergence of new digital technology created a new medium – short video. Compared with news reports and long video, short video is fragmented, intuitive and concise (Li and Ji, 2020; Wu et al., 2021).

The short video platform economy has entered a stage of rapid development recently, at the same time as the outbreak of the COVID-19 epidemic around the world (QuestMobile, 2020b). New short video platforms keep emerging, such as the entertainment and social short video platforms "Kuai shou", "Dou yin", "Little Red Book", "bilibili" and "Tencent video". According to the 50th "China Statistical Report on Internet Development" released by the China Internet Network Information Center (CNNIC), as of June 2022, the number of short video users in China reached 962 million, an increase of 28.05 million compared with 2021, accounting for 91.5% of the total number of Internet users" with "the number of short video users in China accounted for 91.5% of the total number of Internet users, an increase of 28.05 million from 2021, reaching 962 million (CNNIC, 2022). And, short videos attract users of different ages (QuestMobile, 2020a).

As an emerging form of entertainment, a short video platform constantly stimulates users to accept new content and entertainment (Nam and Jung, 2021). Users can get into a flow experience by using short video platforms. Flow experience as a spiritual reward entices them to continue to watch short videos, so users' use may develop into problematic use (Huang et al., 2022). For example, the use of short video platforms will lead to user distraction, poor time management and a waste of time that could be used for learning (Hong et al., 2014; Gao et al., 2017), thus forming a barrier in users' life (Zhang et al., 2019). At the same time, users also experience some physical and mental health problems, such as anxiety and depression (Kitazawa et al., 2018; Fumero et al., 2018). Especially in the context of the COVID-19 epidemic, users with problematic Internet use may develop more serious mental illness (Gecaite-Stonciene et al., 2021). The outbreak of the COVID-19 epidemic has had a considerable impact on people's mental health, and many people feel fear and anxiety (WHO, 2020). Furthermore, short video addiction can negatively affect students' learning (Ye et al., 2022). Short video platforms also use recommendation algorithms to effectively and timely push content of interest to users, which improves users' perceived value and flow experience of short video applications (Cong et al., 2018). However, short video recommendation algorithms also have negative effects, such as Information Cocoons (Meral, 2021). Therefore, short video platforms should pay more attention to the user's resistance to the recommendation algorithm (Velkova and Kaun, 2021).

The user's online flowing experience increases the user's satisfaction and future usage intentions (Hausman and Siekpe, 2009). However, if the user has a negative experience with a product or service, it will lead to the user's regret experience and reduced satisfaction, which will lead to their change products or services (Kang et al., 2009; Bui et al., 2011). Users have different experiences of regret when using different products, but users who overuse social platforms tend to experience more regrets (Kaur et al., 2016). Similarly, users are increasingly relying on short video platforms and are increasingly immersed in watching short videos. Many scholars have studied how these videos increase user activity and immersion, how to produce content, commercial marketing, system management, how to optimize recommendation algorithms, video technology, the flow experience of using short video platforms, etc. (Davis, 2019; Huang et al., 2022; Jacob et al., 2020). Other scholars have studied the negative effects of short video platforms on users. For example, teenagers' addiction to short video platforms and the use of short video platforms by family members will aggravate the degree of addiction among teenagers (Lu et al., 2022). The flow experience of watching short videos has an unfavorable effect on students' learning motivation and happiness (Ye et al., 2022). As one of the short video platforms, Douyin users' flow experience has a momentous effect on addictive behaviors (Qin et al., 2022). Similarly, short video platforms have an impact on the mental health of college students (Wen and Wei, 2022). However, few scholars have studied in depth the negative emotions generated by the negative effects of short video platforms on users, and whether these negative emotions affect users' intention to continue using short video platforms. Therefore, we try to answer the following questions:

RQ1: What are the negative effects of short video platforms on users?

RQ2: How do the negative effects brought about by users using short video platforms, affect their emotions?

RQ3: How does the negative emotion that short video platforms bring to users, affect users' intention to continue using short video platforms?

To address the above questions, this study proposes a preliminary research model to explain the relationship between flow experience, illusion of control and negative emotions by combining flow theory and illusion of control theory. This research mainly focuses on what users think and feel after using short video applications, not the impact of short video content itself on users. Finally, by studying the user's feeling of using the short video platform, it is extended to the user's continued intention to use the short video platform. This study adopts the sequential mixed research method (Venkatesh, et al., 2013; Venkatesh, et al., 2016). The study expands the breadth and depth of research on short videos and enriches the literature related to short video.

The rest of this article is outlined below. Section 2 provides a literature review of flow theory and illusion of control theory. Section 3 proposes a research model based on the literature. Section 4 describes the mixed research method used, first conducting quantitative analysis to draw inferences, then conducting qualitative research to verify the quantitative analysis inferences and supplement the incomplete quantitative analysis. Section 5 summarizes the main findings of the quantitative and qualitative research with a discussion. Section 6 advocates the theoretical implications, practical implications and limitations of this study as well as suggestions for future research.

## 2. Theoretical background and research model

### 2.1. Flow theory

The flow theory originated from Csikszentmihalyi's research on games (Csikszentmihalyi, 1975b). Csikszentmihalyi (1975a) first proposed the theory of immersion. Flow experience is one in which an individual is totally immersed in an activity, ignoring the overall state of everything else. The feeling that an individual is entirely concentrated on an activity is called immersion, also known as flow experience (Csikszentmihalyi and LeFevre, 1989; Tuncer, 2021). Concentration and enjoyment are two main components of flow experience (Ghani and Deshpande, 1994). Flow experience includes not only subjective personal experiences characterized by concentration and enjoyment, but also exploratory and entertaining experiences (Huang and Liao, 2017). Flow experience is characterized by heightened concentration, gradual loss of self-awareness, and a convergence of action and consciousness, accompanied by a strong sense of control and distorted sense of time, a sense of relief and a high level of intrinsic reward (Csikszentmihalyi, 1990).

With the expansion of flow theory, flow theory extends to the field of games and human–computer interaction. Immersion is first derived from the research on the pleasure experienced by players in real games. Moreover, Hoffman and Novak (1996) applied the concept of immersion to searching for information on a network for the first time and introduced the flow theory into the networks and virtual worlds, extending the flow theory to the field of human–computer interaction. Immersion will enhance people's sense of control over the interaction and make people have a positive subjective experience (Hoffman and Novak, 1996). Flow experience is the user's perception during the interaction with the computer and constantly affects the behavior of Internet users.

In previous academic studies, flow experience has often been identified as an optimal experience, that can lead users to have a opportune impression about a brand or software, resulting in longer visits (Esteban-Millat et al., 2014a; Koufaris, 2002; Yang et al., 2014). Therefore, most research is about how to improve users' flow experience to achieve more positive results (Yang et al., 2014). Although flow is usually described as having positive effects, flow is considered negative when someone is doing something unrelated to the task, which means flow is considered to have negative effects during problematic Internet use of entertainment information technology (Argiropoulou and Vlachopanou, 2021). Kaur et al. (2016) have studied the flow experience of online regret. Collins et al. (2009) studied the interindividual and intra - individual impact of flow experience on emotion and showed that high quality flow experience was negatively correlated with low arousal of sadness and disappointment. In addition, Dhir et al. (2016) based on flow experience found that many Facebook users tend to feel more regret.

### 2.2. Illusion of control theory

Langer (1975) first proposed the illusion of control theory. In the psychological literature, illusion of control is widely recognized as an egocentric bias, including overconfidence and unrealistic optimism (DeBondt and Thaler, 1994; Fischhoff et al., 1977; Weinstein, 1980). Also, illusion of control often reinforces skills or abilities by mistaken perceptions of environmental stimuli, even if the outcome is purely random (Fellner, 2009; Yu and Fu, 2019). Therefore, the more controlling individuals are, the more likely they are to overestimate their chances of winning (Burger and Cooper, 1979).

Illusions of control are commonly used in experiments in which subjects are resulted in falsely believe that they can exert control over certain outcomes (Stefan and David, 2013). Langer and almost all the other researchers that have examined the illusion of control have employed statistical tests, but they cannot assess whether a person's or a group's expectations of individual success are too high relative to objective criteria (Langer, 1975; Langer and Roth, 1975).The application of illusion of control theory was extended from uncontrollable to controllable situations (Thompson et al. (2007)).

The illusion of control theory is not only applied to gambling and games (Fu and Yu, 2015; Meng and Leary, 2021; Yu and Fu,2019), it is also applied in other aspects. Martinez et al. (2011) studied the relationship between the illusion of control and risk taking. Bandera et al.

(2018) studied the contradictory relationship between entrepreneurial self-efficacy traits and the illusion of control. In addition, external measures can enhance individuals' perception of control (Studer et al., 2020), and the illusion of control may also result in people's overestimation of authority (Sloof and von Siemens, 2017).

### 2.3. Flow experience and negative emotions

#### 2.3.1. Flow experience and low efficiency

In relation to online immersion, Trevino and Webster (1992) found that flow experience is one of the key factors. This was identified by studying the influence of computer-mediated communication technology on work efficiency. If the immersion is not related to an employee's task, it may lead to a decrease in the efficiency at work (Sharafi et al., 2006). Therefore, in work or study, high user participation and low interference can reduce the occurrence of regrets (Kuhnle and Sinclair, 2011), on the contrary, low efficiency brought by social media may cause users to have more regrets (Dhir et al.,2016). When people receive a quantity of fragmented information in a short period of time, their concentration will be reduced and it will be difficult for people to concentrate for a long time for deep thinking (Hong et al., 2014), which will lead to lower efficiency in daily learning and work. Therefore, we propose the following hypothesis:

H1: Users' inefficiency in work or study due to being immersed in short video platforms, can lead them to have negative emotions.

#### 2.3.2. Flow experience and time distortion

Online immersion weakens users' perception of time, leading to time distortion. Time distortion is when the individual is unaware of the passage of time, so the individual feels that time passes quickly after leaving the immersion state (Novak et al., 2000). When individuals are in flow experience, their perceived passing of time, loses consistency with actual elapsed time (Esteban-Millat et al., 2014b). Thatcher et al. (2008) found that problematic Internet use, online procrastination and immersion are all due to lack of control over online time. Similarly, Kaur et al. (2016) found that immersion on Facebook increased the level of regret experienced by users. When users use short video apps, they often feel that time passes quickly (Wu et al., 2021), resulting in a misalignment between their perceived time changes and the actual time spent on short video platform, even gradually blurring the concept of time. Therefore, we propose the following hypothesis:

H2: Users' time distortion due to being immersed in short video platforms can lead them to have negative emotions.

#### 2.3.3. Flow experience and harm to health

Online immersion will do harm to users' physical and mental health. Problematic Internet use can leave people, especially teenagers, with problems such as dysfunctional relationships, lost social skills, and impaired physical and mental health (Chao and Yu, 2021; Öztürk and Özmen, 2016). If users are not restrained when watching short videos, short videos will continuously stimulate the user's vision and hearing to keep the user in a state of excitement for a long time, which may damage to users' physical and mental health (Wu et al., 2021). Therefore, we propose the following hypothesis:

H3: Users' physical and mental health damage due to being immersed in short video platforms, can lead them to have negative emotions.

#### 2.3.4. Flow experience and addiction

Online immersion can exacerbate addictive behaviors. Users who stay immersed in an experience for prolonged periods are more likely to develop addictive behaviors (Salehan and Negahban, 2013). Internet addiction can lead individuals to have anxiety, depression and other psychological problems (Fumero et al., 2018; Kitazawa et al., 2018). Social interaction is the main reason that people become addicted to watching short videos, and it is more important than app dependency (Zhang et al., 2019). Furthermore, addiction has more negative effects on mood than positive effects (Błachnio et al., 2017; Tang et al., 2016). Flow experience has a significant impact on short video addictive behavior (Qin et al., 2022). Short video apps are a double-edged sword, and fair use can relieve stress, but overuse can lead to addictive symptoms. Therefore, users may not be able to control themselves because of excessive immersion in short videos, resulting in addictive behaviors. Thus, we propose the following hypothesis:

H4: Users' addictive behavior due to being immersed in short video platforms, can lead them to have negative emotions.

#### 2.3.5. Flow experience and online procrastination

Online immersion can lead to online procrastination. Procrastinators may prefer to use the Internet rather than study because it is more interesting and enjoyable to use it in flow experience (Thatcher et al. 2008). Therefore, online procrastination refers to using the Internet to escape other boring, unchallenging, or disheartening tasks (Thatcher et al., 2008). It is also as known as computer procrastination (Breems and Basden, 2014). When a person is more inclined to procrastinate online, he or she is more likely to suffer negative consequences (Hernández et al., 2019). Procrastination can also lead to negative emotions such as depression, anxiety, anger, worry, remorse, and shame (Häfner et al., 2014; Pychyl et al., 2000; Rothblum et al., 1986). Users can get instant happiness by watching short videos, and this effortless way of getting happiness may cause users to procrastinate when faced with tasks that require deep thinking or tasks that are difficult. Therefore, we propose the following hypothesis:

H5: Users' online procrastination due to being immersed in short video platforms, can lead them to have negative emotions.

### 2.4. Illusion of control and negative emotions

Illusions of control are people's tendency to believe an event they want, and when this outcome happens independently of their behavior, they mistakenly believe that they are controlling it happening (Matute et al., 2007). This delusion is often at the heart of superstitious and pseudoscientific thinking (Matute et al., 2007; Matute and Blanco, 2014). The existence of this effect is not surprising. Throughout human history, humans have believed that they control the occurrence of certain events. Now, people even can control the results on the Internet (Blanco et al., 2009).

In most cases, illusion of control will bring negative effects. We infer that users' illusion of control over the use of short video applications will cause users to have negative emotions, thereby affecting users' intention to continue using them. People tend to overestimate their ability to manage themselves. When users find that they spend too much time watching short videos or it affects their own lives, they may take a series of measures to change their habit of watching short videos. However, if users cannot control themselves, they may have negative emotions and fall into a vicious circle of illusion of control. Therefore, we propose the following hypothesis:

H6: Users' illusion of control over the use of short video applications, can cause them to have negative emotions.

### 2.5. Negative emotions and the intention to continue using short video platforms

Emotions not only form beliefs and attitudes but also help guide decision-making (Lazarus and Folkman, 1984). According to previous literature, in the research on technology usage and persistence, scholars mainly focus on positive effects and positive emotions, and find that positive emotions can increase users' intention to use (Beaudry and Pinsonneault, 2010; Turel, 2015). However, there is little research on the effect of negative emotions on users' continued intention to use (Lee, 2016). Emotions represent the mental state of taking action, so individual actions are often influenced by emotions (Lazarus, 1991). In the research on users' use of short video platforms, the negative emotions that this study mainly focuses on are regret, anxiety and sadness (Tarafdar et al., 2010). Combined with the above flow experience and illusion of control, the related effects and negative emotions of users, we can speculate that users of short video platforms may also have some negative emotions because of using short video platforms, which in turn will affect their intention to continue using short video platforms. Therefore, we propose the following hypothesis:

H7: Negative emotions of users reduce their intention to continue using short video apps.

Based on flow theory and illusion of control theory, this study proposes a model that takes users' use of short video platforms as an example to explain what factors cause users to have negative emotions after using short video platforms, and how these negative emotions affect their intention to continue using short video platforms. Our research model is shown in Fig. 1.

## 3. Research methodology

In this study, in order to achieve scientific rigor, we used a mixed research approach, combining quantitative and qualitative research to synthesize findings (Creswell, 2003; Venkatesh et al., 2013). Mixed methods research combines the design advantages of qualitative and quantitative research and can explain various phenomena more comprehensively and rationally (Venkatesh et al., 2013; Venkatesh et al., 2016). Mixed methods design yield richer and more robust conclusions than single methods (Venkatesh et al., 2016) and are more helpful in addressing confirmatory and experimental research questions (Teddlie and Tashakkori, 2003; Teddlie and Tashakkori, 2009).

Quantitative data were first collected and analyzed (Creswell, 2003). Second, qualitative data were collected to validate conclusions drawn from quantitative data and to complement the incomplete analysis of quantitative data (Plano Clark and Ivankova, 2015). Quantitative data and qualitative data are collected sequentially, complementing each other, and correlating with each other (Plano Clark and Ivankova, 2015). Finally, we determine the boundary conditions for meta-reasoning by analyzing the complementarity and contradiction of qualitative data with quantitative conclusions (Cheng et al., 2022; Lewis and Grimes 1999; Venkatesh et al. 2013; Srivastava and Chandra, 2018). The whole process of the method is shown in Fig. 2.

### 3.1. Quantitative study

#### 3.1.1. Quantitative data collection

*3.1.1.1. Questionnaire.* All questionnaire items in this study were scored on a five-point Likert scale starting from 1 (indicating "strongly disagree") to 5 (indicating "strongly agree"). The items and their sources are presented in Appendix A.

The final version of the questionnaire was distributed via the Internet. We used a snowball sampling strategy to send links to respondents through social media such as QQ and WeChat. Specifically, we adopted a bonus system, setting up cash red envelopes in the link, and the amount of red envelopes ranges from 1 to 5 RMB. When the respondents complete the answers, they can draw 100 % of the red envelopes. Then, we conducted a valid questionnaire review in the background, and the valid questionnaire would be rewarded with red envelopes. The red envelope reward mechanism can effectively encourage respondents to share links through WeChat Moments, QQ Space, or send them to friends and family. The structure of the questionnaire is as follows.

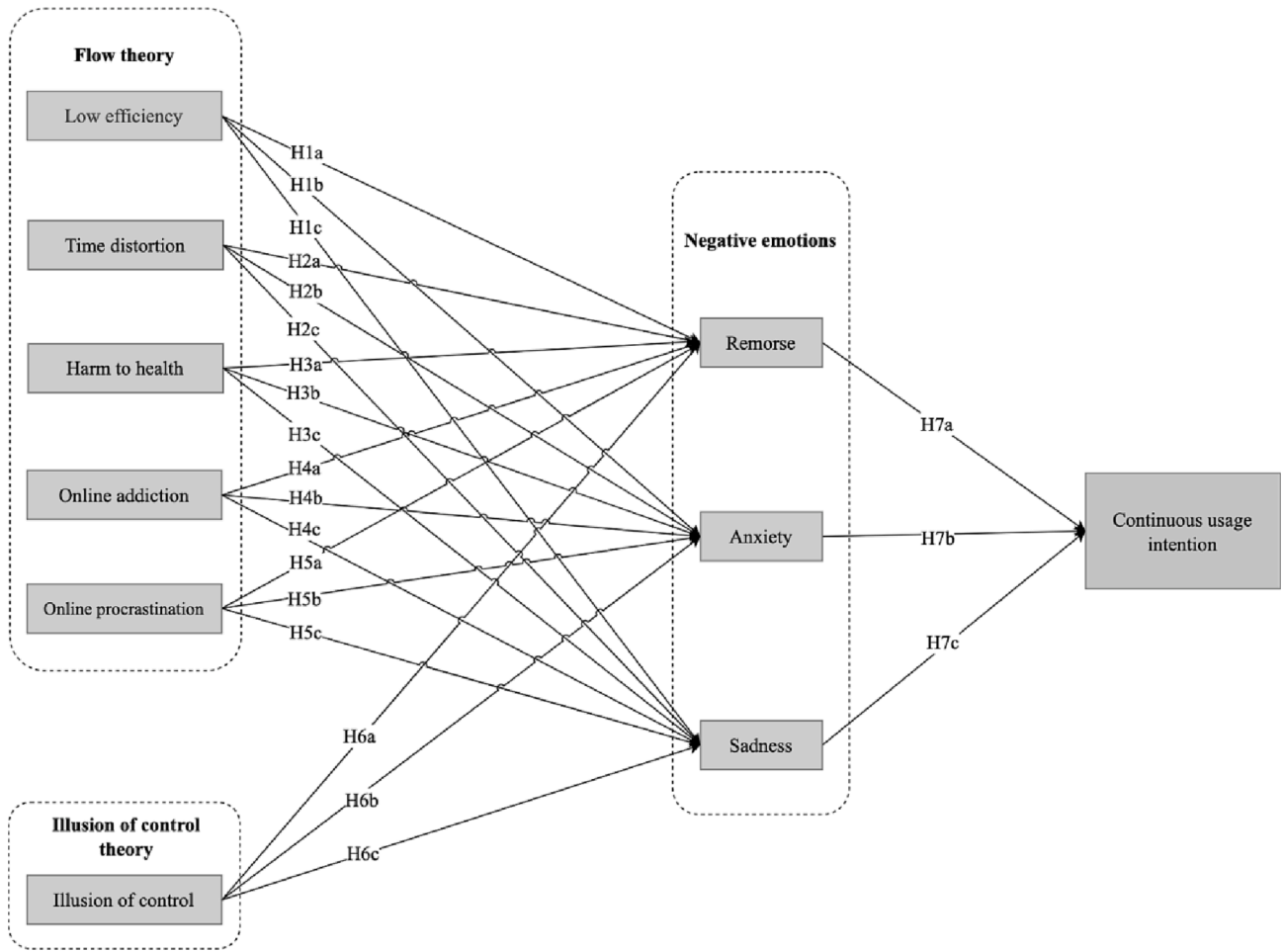

**Fig. 1.** Research Model.

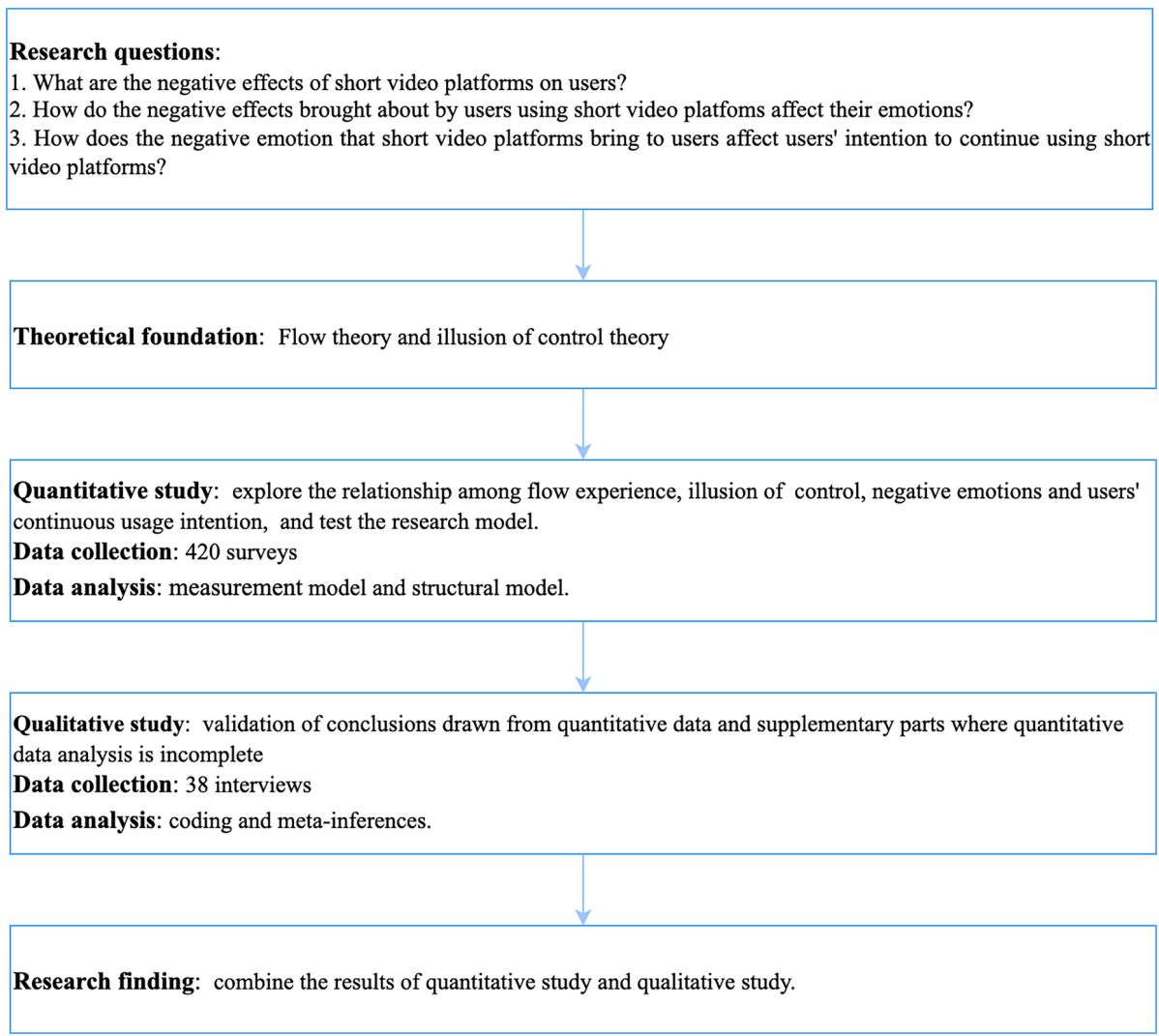

**Fig. 2.** Mixed-methods approach.

*3.1.1.1.1. Low efficiency.* The measure has been adapted from the LEAPS (The Lam Employment Absence and Productivity Scale) concerning low efficiency (Lam et al., 2009). It consisted of four items addressing low efficiency caused due to short video applications use. In this study, the complete scale showed a Cronbach's alpha of 0.88.

*3.1.1.1.2. Time distortion.* The time distortion scale was adapted from similar questionnaires which measure time distortion (Koufaris, 2002; Novak et al., 2000). The questionnaire consists of 3 items and measures the extent in which an individual feels time passing while using short video applications. In this study, the complete scale showed a Cronbach's alpha of 0.847.

*3.1.1.1.3. Harm to health.* The Chinese Psychosomatic Health Scale (CPSHS) has a total of 134 questions, which was compiled by Liyi Zhang, chief doctor of the Chinese People's Liberation Army of the People's Republic of China, and jointly developed with 27 provincial and municipal fraternal units across the country (Zhang et al., 1993). The questionnaire consisted of four items addressing harm to health caused due to short video applications use. In this study, the complete scale showed a Cronbach's alpha of 0.862.

*3.1.1.1.4. Online addictiveness.* Short video application addiction measures three criteria for internet addiction, a scale adapted from Young (1998). By collecting respondents' responses to Internet addiction, this study effectively summarizes the actual situation of people's addiction to short video applications (Yang et al., 2014; Yang and Tung, 2007). The questionnaire consists of 3 items. In this study, the complete scale showed a Cronbach's alpha of 0.822.

*3.1.1.1.5. Online procrastination.* To measure procrastination caused by addicted viewing of short videos, we refer to literature such as online procrastination caused by problematic Internet use (Thatcher et al., 2008; Hernández et al., 2019). The questionnaire consists of 3 items. In this study, the complete scale showed a Cronbach's alpha of 0.875.

*3.1.1.1.6. Illusion of control.* The illusion of control section used the Illusion of Control Beliefs scale (ICB) to test participants' degree of control over viewing a short video (Moore and Ohtsuka, 1999). It also applies to the assessment of Chinese youth's belief in ICB (Fu and Yu 2013). In this study, the complete scale showed a Cronbach's alpha of 0.866.

*3.1.1.1.7. Remorse and sadness.* The scale of user's remorse and sadness emotion after using short video applications was adapted from Positive and Negative Affect Schedule (PANAS) (Watson et al., 1988) and four criteria of remorse and three criteria of sadness were measured separately. According to previous studies, the scale has good reliability and validity and has a wide range of applications. In this study, the remorse scale showed a Cronbach's alpha of 0.928 and the sadness scale showed a Cronbach's alpha of 0.869.

*3.1.1.1.8. Anxiety.* The scale of user's anxiety emotion after using short video applications was adapted from Self-Rating Anxiety Scale (SAS) (Zung (1965)) and measured four criteria of anxiety. SAS is a psychological scale for assessing anxiety and is used by psychologists and psychiatrists as a tool to measure the severity of anxiety during the treatment of patients. In this study, the complete scale showed a Cronbach's alpha of 0.909.

*3.1.1.1.9. Continuous usage intention.* The intention to continue to use is the intention of users to continue to use the Information System after having used (Bhattacherjee, 2001). According to the Information System continuous use intention measurement scale designed, this scale has compiled the following three measurement items related to users' intention to continue using short video platforms. In this study, the complete scale showed a Cronbach's alpha of 0.890.

*3.1.1.2. Pilot study.* After the questionnaire design is completed, this research's questionnaire is validated in two steps. In order to improve the validity of the questionnaire, we invited four volunteers to participate in revising and refining the wording of the questionnaire to ensure that the participants could understand the questionnaire items well. A pre-test was conducted, first handing out questionnaires in the class group, collecting a small number of samples (n = 72) for reliability and validity testing, and distributing the questionnaires to a large scale after passing the test.

*3.1.1.3. Participants.* A total of 582 respondents' questionnaire data was collected, and 420 valid responses were obtained. Detailed demographic information of the sample is shown in Table 1. Among all the respondents, 47.38 % were male and 52.62 % were female, and the gender ratio is relatively balanced. A majority of them were young people aged between 18 and 24 (56.43 %), and most of the respondents had received college education (76.43 %). Furthermore, most of the respondents watch short videos for>0.5 (84.52 %) hours per day. More than half of the respondents watch short videos every day for 0.5 h-2 h (51.43 %). The percentage of users watching short videos 0 h–1 h (15.48 %), 2 h–3 h (16.9 %) and the >3 h (16.19 %) is nearly equal.

*3.1.2. Quantitative data analysis*

In this study, the virtual research model was explored using structural equation modeling (SEM) with maximum likelihood estimation. SEM, which includes explicit and latent variables, is a statistical method for founding, estimating, and testing relationships between variables. Specifically, we attempt to explain the relationship between variables based on covariance-based (CB) SEM (Hair et al., 2017). Therefore, a measurement model and a structural model by SPSS v.26 and Amos v.26 for statistical analysis using covariance CB-SEM was applied. Specifically, SPSS v.26 was used for validity analysis, and Amos v.26 was used for reliability and common method bias. What's more, the sample size collected in this study also met the requirement of covariance CB-SEM (>200) (Hair et al., 2012). Before the analysis, a check was made on the normality of the data distribution. All the items had skewness and kurtosis values <3. This suggests that these data are suitable for Amos' analysis.

*3.1.2.1. Common method bias.* Since the single data collection method of questionnaire is used in this research, it will inevitably lead to high correlation between variables, that is, there may be common method bias (CMB) (Campbell and Fiske, 1959). CMB is considered a procedural and statistical remedy (Podsakoff et al., 2012). Therefore, we examined whether the data were affected by CMB in terms of the questionnaire process and diagnostic test CMB. First, when we distribute the questionnaire online, we assure respondents that they will participate anonymously to encourage them to answer the questionnaire questions truthfully. Second, this paper uses the common marker variable method to test the data (Lindell and Whitney, 2001). This paper uses age as the label variable, because the label variable should be variable with the smallest correlation with the real variable (Williams et al., 2010). The correlation coefficient between the age and the real variable is 0, indicating that the age is not correlated with the real variable, which is a good label variable. This paper adopts the confirmatory factor analysis (CFA) method to construct the baseline model and the control model with the label variable added, the difference between the chi-square value of the baseline model of 27.810 (df = 149) and the chi-square value of the control model of 25.295 (df = 159) is 2.515 (df = 10), which does not reach a significant level (p > 0.05). This suggests that there is no substantial CMB in the data studied in this paper (Podsakoff et al., 2003).

*3.1.2.2. Measurement model.* In this study, the measurement characteristics of the instrument were first evaluated in the data analysis, and then the structural relationships of the model were examined (Anderson and Gerbing 1988; Hair et al. 2019a). Furthermore, in order to evaluate the measurement model, three types of validity were tested: content validity, convergent validity, and discriminant validity. In this study, measures used to assess the effectiveness of content adequately covered the model (Straub et al., 2004). Therefore, in the questionnaire design stage, the validity of the content was checked by measuring the consistency with the existing literature (Srivastava and Chandra, 2018).

Firstly, the reliability of the scale is verified. As shown in Table 2, Cronbach's alpha values are above 0.8 and greater than the threshold value of 0.7, indicating the reliability of the model is acceptable (Nunnally, 1967). In this study, multicollinearity was assessed by calculating the variance inflation factor (VIF), with all VIF values below the recommended threshold of 10 (Hair et al., 2019b).

Secondly, the convergent validity of the multiple-items is tested. Loadings of factors are above 0.7 (Fornell and Larcker, 1981) and all the Average Variance Extracted (AVE) are above 0.6 and exceeded the required value 0.5, indicating that convergence validity is satisfied (Bagozzi and Yi, 1988).

Finally, the discriminant validity is evaluated. The square root of AVE is greater than the correlation coefficient of each latent variable as shown in Table 3, which satisfies the discriminant validity (Fornell and Larcker, 1981)

The goodness-of-fit index used in this study, such as the Comparative Fit Index (CFI), Tucker – Lewis index (TLI), Incremental Fit Index (IFI), Root Mean Square Error of Approximation (RMSEA), and Standardized Root Mean-squared Residua (SRMR). As shown in Table 4, the fitted index of the structural equation model in this study is as follows: The chi-square DF ratio $x^2/df$ is 2.612 (its degree of freedom should be below 3), RMSEA is 0.062 (0.08 means good fit, and 0.05~0.08 means quite good fit), SRMR is 0.074 (0.08 means good fit, and 0.05~0.08 means quite good fitting), CFI is 0.907 (≥0.9), TLI is 0.900 (≥0.9), and IFI is 0.908 (≥0.9). (Hu and Bentler, 1999; Westland, 2019).

**Table 1**
Demographics of respondents (N = 420).

| Category | Item | Frequency | Percent (%) |
|---|---|---|---|
| Gender | Male | 199 | 47.38 |
| | Female | 221 | 52.62 |
| Age | <18 | 13 | 3.10 |
| | 18–24 | 237 | 56.43 |
| | 25–34 | 130 | 30.95 |
| | 35–54 | 37 | 8.81 |
| | >=55 | 3 | 0.71 |
| Education | Senior high school and below | 28 | 6.67 |
| | Associate degree | 71 | 16.90 |
| | Bachelor's degree | 234 | 55.72 |
| | Master's degree | 73 | 17.38 |
| | Doctoral degree | 14 | 3.33 |
| Usage time everyday | 0–0.5 h | 65 | 15.48 |
| | 0.5–1 h | 101 | 24.05 |
| | 1–2 h | 115 | 27.38 |
| | 2–3 h | 71 | 16.90 |
| | >3 h | 68 | 16.19 |

**Table 2**
Confirmatory factor analysis for measurement model.

| Construct | Items | Factor Loading | Cronbach's α | CR | AVE |
|---|---|---|---|---|---|
| Low efficiency (LE) | LE1 | 0.855 | 0.880 | 0.888 | 0.666 |
|  | LE2 | 0.855 |  |  |  |
|  | LE3 | 0.835 |  |  |  |
|  | LE4 | 0.709 |  |  |  |
| Time distortion (TD) | TD1 | 0.849 | 0.847 | 0.847 | 0.649 |
|  | TD2 | 0.782 |  |  |  |
|  | TD3 | 0.783 |  |  |  |
| Harm to health (HH) | HH1 | 0.714 | 0.862 | 0.881 | 0.650 |
|  | HH2 | 0.810 |  |  |  |
|  | HH3 | 0.868 |  |  |  |
|  | HH4 | 0.825 |  |  |  |
| Online addiction (OA) | OA1 | 0.770 | 0.822 | 0.831 | 0.622 |
|  | OA2 | 0.844 |  |  |  |
|  | OA3 | 0.749 |  |  |  |
| Online procrastination (OP) | OP1 | 0.831 | 0.875 | 0.887 | 0.724 |
|  | OP2 | 0.853 |  |  |  |
|  | OP3 | 0.829 |  |  |  |
| Illusion of control (IC) | IC1 | 0.869 | 0.866 | 0.865 | 0.762 |
|  | IC2 | 0.877 |  |  |  |
| Remorse (R) | R1 | 0.844 | 0.928 | 0.934 | 0.780 |
|  | R2 | 0.919 |  |  |  |
|  | R3 | 0.905 |  |  |  |
|  | R4 | 0.863 |  |  |  |
| Anxiety (A) | A1 | 0.896 | 0.909 | 0.907 | 0.711 |
|  | A2 | 0.910 |  |  |  |
|  | A3 | 0.826 |  |  |  |
|  | A4 | 0.730 |  |  |  |
| Sadness (S) | S1 | 0.708 | 0.869 | 0.876 | 0.704 |
|  | S2 | 0.914 |  |  |  |
|  | S3 | 0.881 |  |  |  |
| Continuous usage intention (CUI) | CUI1 | 0.869 | 0.890 | 0.891 | 0.731 |
|  | CUI2 | 0.883 |  |  |  |
|  | CUI3 | 0.812 |  |  |  |

*3.1.2.3. Structural model.* The estimated results are presented in Tables 5a and 5b and Fig. 3. This study hypothesizes the relationship between the flow experience, illusion of control and users' negative emotions, illusion of control and users' negative emotions, as well as between users' negative emotions and their continuous usage intention in short video apps. According to the result, 59 % of the variance in continuous usage intention is from short video apps. 42 %, 40 % and 36 % of variances are explained by remorse, anxiety and sadness. Furthermore, the percentages of variances were all >30 %, which suggested that it was a satisfactory model (Falk and Miller, 1992).

This paper tested the hypothesized relationship between flow experience and negative emotions (i.e., remorse, anxiety and sadness). According to the research results, flow experience of low efficiency is positively related to remorse (β = 0.125, p < 0.05). Flow experience of low efficiency is not significantly related to anxiety (β = 0.060, p > 0.05) and sadness (β = −0.087, p > 0.05). Flow experience of time distortion is positively related to remorse (β = 0.213, p < 0.01). Flow experience of time distortion is not significantly related to anxiety (β = 0.030, p > 0.05) and sadness (β = 0.021, p > 0.05). Flow experience of harm to health is positively related to all three, remorse (β = 0.185, p < 0.01), anxiety (β = 0.478, p < 0.001) and sadness (β = 0.383, p < 0.001). Also, flow experience of online addiction is related to all three, remorse (β = 0.198, p < 0.01), anxiety (β = 0.232, p < 0.001) and sadness (β = 0.476, p < 0.001). Flow experience of online procrastination is positively related to remorse (β = 0.149, p < 0.05) and anxiety (β = 0.145, p < 0.05). Flow experience of online procrastination is not significantly related to sadness (β = 0.064, p > 0.05). Therefore, H1a-H5a, H3b-H5b, H3c-H4c are supported, H1b-H2b, H1-H2c, H5c are rejected. In

**Table 4**
Model fit.

| Indicator | Value | Standard |
|---|---|---|
| $x^2$ | 1421.06 | The smaller the better |
| $x^2/df$ | 2.612 | ≤3 |
| IFI | 0.908 | ≥0.9 |
| CFI | 0.907 | ≥0.9 |
| TLI | 0.9 | ≥0.9 |
| RMSEA | 0.062 | ≤0.08 |
| SRMR | 0.074 | ≤0.08 |

**Table 5a**
Effect overview.

| Effect | Standardized weight | p value | Result |
|---|---|---|---|
| LE→R | 0.125 | 0.049* | Supported |
| LE→A | 0.060 | 0.362 | Rejected |
| LE→S | −0.087 | 0.180 | Rejected |
| TD→R | 0.213 | 0.006** | Supported |
| TD→A | 0.030 | 0.707 | Rejected |
| TD→S | 0.021 | 0.792 | Rejected |
| HH→R | 0.185 | 0.003** | Supported |
| HH→A | 0.478 | *** | Supported |
| HH→S | 0.383 | *** | Supported |
| OA→R | 0.198 | 0.002** | Supported |
| OA→A | 0.232 | *** | Supported |
| OA→S | 0.476 | *** | Supported |
| OP→R | 0.149 | 0.020* | Supported |
| OP→A | 0.145 | 0.026* | Supported |
| OP→S | 0.064 | 0.318 | Rejected |
| IC→R | 0.065 | 0.182 | Rejected |
| IC→A | 0.105 | 0.033* | Supported |
| IC→S | 0.136 | 0.005** | Supported |
| R→CUI | −0.292 | *** | Supported |
| A→CUI | 0.049 | 0.404 | Rejected |
| S→CUI | 0.412 | *** | Supported |
| Age→CUI | 0.257 | *** | Supported |
| Education→CUI | −0.115 | 0.016* | Supported |
| UsageTime→CUI | 0.198 | *** | Supported |

**Table 3**
Discriminant validity results (Note: the diagonal values are the square root if AVE.).

| Constructs | IC | OP | OA | HH | LE | TD | S | A | R | CUI |
|---|---|---|---|---|---|---|---|---|---|---|
| IC | **0.873** |  |  |  |  |  |  |  |  |  |
| OP | −0.269 | **0.851** |  |  |  |  |  |  |  |  |
| OA | 0.217 | 0.618 | **0.789** |  |  |  |  |  |  |  |
| HH | 0.034 | 0.438 | 0.490 | **0.806** |  |  |  |  |  |  |
| LE | −0.059 | 0.375 | 0.272 | 0.505 | **0.816** |  |  |  |  |  |
| TD | −0.136 | 0.344 | 0.174 | 0.418 | 0.549 | **0.806** |  |  |  |  |
| S | 0.022 | 0.458 | 0.647 | 0.594 | 0.260 | 0.211 | **0.839** |  |  |  |
| A | 0.020 | 0.475 | 0.528 | 0.654 | 0.395 | 0.308 | 0.495 | **0.843** |  |  |
| R | −0.046 | 0.453 | 0.449 | 0.491 | 0.416 | 0.390 | 0.388 | 0.409 | **0.883** |  |
| CUI | 0.022 | 0.087 | 0.166 | 0.140 | 0.014 | −0.003 | 0.321 | 0.139 | −0.088 | **0.855** |

**Table 5b**
Selected quotes from interviewees.

| Category | Construct | Selected quote |
|---|---|---|
| Flow experience | Low efficiency | "Watching short videos will reduce my usual study time and make it difficult for me to concentrate on one thing, so I always want to watch short videos when I study." (A1) |
| | Time distortion | "I feel that the sense of satisfaction and happiness is particularly easy to obtain when I watch short videos, which will cause me to resist difficult study or work tasks. Work interest has diminished." (A27) "I usually watch two hours of short videos after I get home from get off work, and it does feel like the time flies by quickly." (A6) |
| | Harm to health | "Because when you watch short videos, you only need to swipe up and down and the short video platform will recommend similar videos, so I often forget the time when I watch." (A10) |
| | Online addiction | "The happy time is always very short. If you don't control the time well when you watch short videos, it will feel more like time is passing." (A14) "I feel lost and empty after watching short videos, and I feel like I wasted a lot of time watching something purely for pleasure." (A10) |
| | Online procrastination | "I would feel very lost and empty if I exceeded the scheduled time and had a considerable impact on my learning. Because short videos can only bring me short-lived joy, it can't fill my heart like learning." (A14) |
| Illusion of control | Illusion of control | "If you watch short videos before going to bed, you may fall asleep, just lie in bed and feel restless and can't fall asleep." (A36) "Watching short videos is now something I have to do every day, and I have formed a habit. If I don't, I will feel uncomfortable." (A1) |
| Negative emotions | Remorse | "I think short video platforms are recommended to users based on the type of users they have previously watched. This situation will make people addicted to short videos." (A34) |
| | Anxiety | "If I encounter learning difficulties and feel that I have enough time, I want to delay learning by watching short videos." (A7) |
| | Sadness | "When the learning task is particularly heavy, I will choose to watch short videos to escape temporarily. Then I would stay up all night to finish the task, but I would feel a sense of guilt." (A16) |
| Users' response to platforms | Continuous usage intention | "I will control myself not to watch short videos when I have something important to do. I control myself to watch short videos every day to 0.5 h-1 h." (A26) "I try to avoid the video recommendation function of various short video apps." (A29) "If I watch the short video for more than three hours, the phone will alarm to remind me that I have watched too long today." (A32) "I feel guilty if I watch short videos for too long." (A18) "After watching the short videos, I feel remorse and think I shouldn't have done it." (A25) "If I can't control myself to watch short videos, I feel guilty and self-blame." (A31) "If I had to do my own thing after watching short videos, I would feel irritable and tired. It may be because too much |

**Table 5b** (*continued*)

| Category | Construct | Selected quote |
|---|---|---|
| | | information is received by the brain after watching it for too long." (A2) "I feel more anxious when I can't complete tasks due to watching short videos." (A21) "When I was watching short videos, my mood would be ups and downs, and after I finished it, I would feel that it was difficult to calm down." (A25) "I'm in a better mood when I watch a short video, but if I think back on it and it's taking up my time, I'm in a low mood." (A1) "After watching short videos that are purely hedonic for a long time, I will feel like wasting my time."(A9) "I will not spend more time on short videos in the future, because now the pressure of work and study is increasing, I have to control the time I spend on short videos." (A27) "I don't think it's a good habit to watch short videos, so I don't want to recommend short video apps to others." (A29) |

addition, this paper tested the hypothesized relationship between illusion of control and negative emotions. The results show that illusion of control is not significantly related to remorse (β = 0.065, p > 0.05). And illusion of control is positively related to anxiety (β = 0.105, p < 0.05) and sadness (β = 0.136, p < 0.01). Consequently, H6a is rejected and H6b, H6c are supported. Finally, this paper investigated the causal relationships between users' negative emotions and their continuous usage intention in short video apps. The results indicate that remorse (β = −0.292, p < 0.001) negatively affects users' continuous usage intention toward short video apps. Sadness (β = 0.412, p < 0.001) is significant and positively affects users' continuous usage intention in short video apps. Anxiety (β = 0.049, p > 0.05) does not significantly affect users' continuous usage intention in short video apps. Thus, H7a and H7c are supported, H7b is rejected. Additionally, age (β = 0.257, p < 0.001), and usage time every day (β = 0.198, p < 0.001), have a positive effect on users' continuous usage intention in short video apps. Education (β = −0.115, p < 0.05) is negatively affecting users' continuous usage intention in short video apps.

### 3.2. Qualitative study

#### 3.2.1. Qualitative data collection

This study collected qualitative data using semi-structured interviews to validate and supplement quantitative analysis conclusions. The interview questions consist of a combination of closed-ended and open-ended questions, and they are split into four main parts. To begin with, respondents described their personal experiences with short video apps and whether their watching of short videos changed after the COVID-19 pandemic outbreak. Second, we asked them to talk about the impact of short videos on their lives and what factors had certain negative effects on them. The third section tried to understand the negative emotions of respondents due to the impact of short videos on them and their intention to continue using short video platforms. Finally, the respondents were asked open-ended questions, these were related to the respondents' satisfaction with the short video platform they are currently using and their suggestions for the platform.

To begin with, we followed a list to ask interview questions, and the interviewers were all trained (Eisenhardt, 1989). The interviewer has the flexibility to react based on the respondent's answer, and after receiving an interesting answer, would further discuss with the respondent to obtain more findings. Second, we contacted respondents through social media such as WeChat and informed them of the basic criteria for interviews. We asked respondents in advance if they have experience with short videos, and informed them about the interview time. Finally, in order to ensure the authenticity of respondents' answers, we promise that all respondents will remain anonymous and keep their information confidential. After obtaining the respondent's permission, we recorded the interviews, and then convert the recordings into text for qualitative analysis.

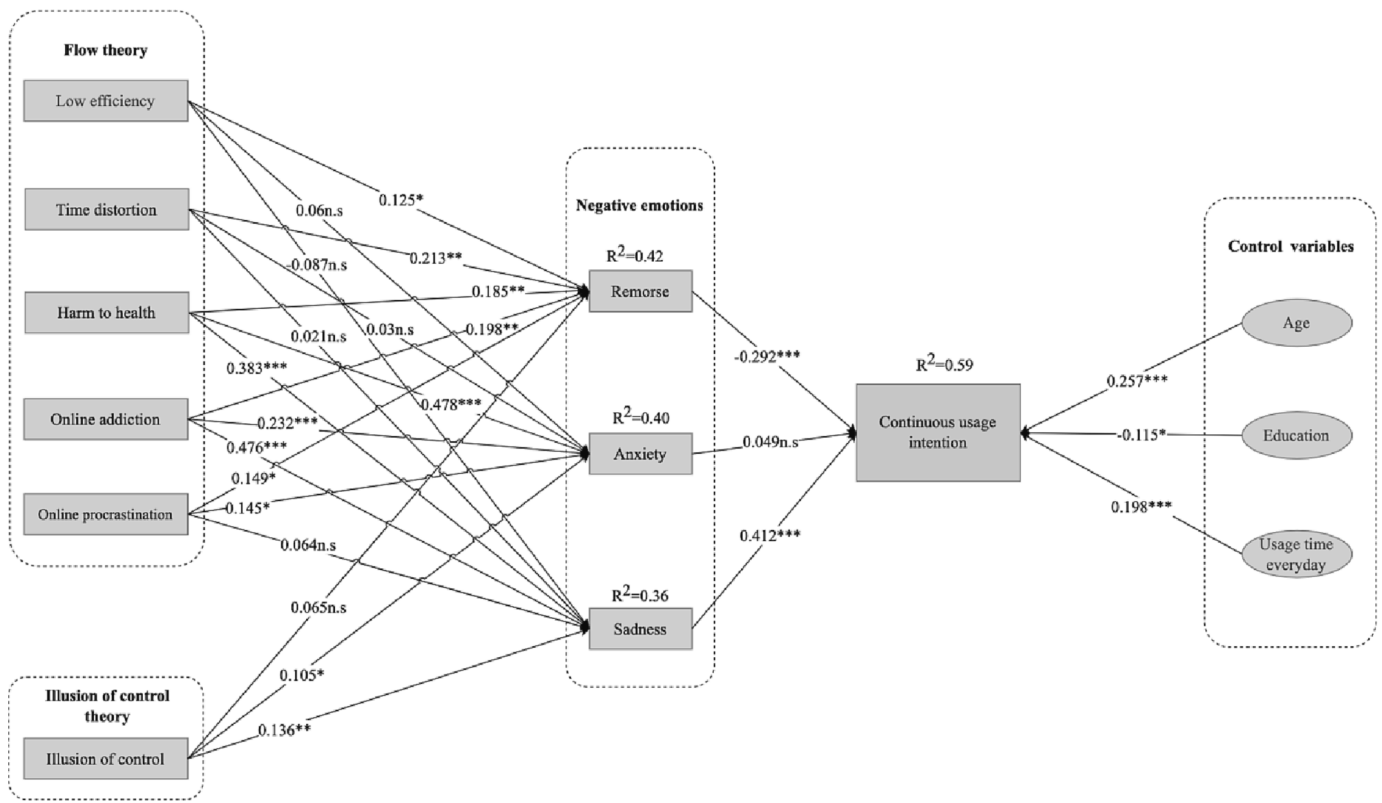

**Fig. 3.** Structural Model Note: ***$p < 0.001$; **$p < 0.01$; *$p < 0.05$; n.s.: not significant.

Over a one-week period, we collected data from a total of 38 interviews, coded A1-A38, with an average interview time of approximately 18 min and a 100 % response rate. Among them, 20 male respondents and 18 female respondents, most of whom are between the ages of 18 and 34, and have a bachelor's degree or above.

*3.2.2. Qualitative data analysis*

The main purpose of qualitative data analysis is to find out the detailed impact mechanism of short videos on users and the user's intention to continue to use short video apps. Qualitative data were coded using induction so that each interview data could be divided into one or more categories. The coding scheme is derived from the research question and the structure of the study, and records responses and non-quantitative analysis factors that differ from the quantitative analysis results, and the entire coding process is iterative to improve generalizability (Srivastava and Chandra, 2018). The main steps of data analysis are as follows:

**Step 1**: Identify phrases and sentences. We marked all the phrases and sentences related to the questions that we were studying by carefully reading each interview transcript.

**Step 2:** Classify data. We group sentences and phrases with similar meanings and use keywords to describe the group content.

**Step 3**: Identify key structures based on theory. Firstly, based on the existing research and theory, the theory is combined with the keywords obtained in step 2 to find the theoretical support corresponding to the keywords. Second, we repeat steps 1 and 2 to summarize and refine the keywords obtained in step 2, and correlate the results with a higher-level theoretical structure.

**Step 4:** Compare keywords and variable names. Refer to the research model variables of quantitative research to modify the keywords summarized in the interview data, compare which expression is more accurate, and finally get the most accurate expression.

*3.2.2.1. Corroboration and confirmation.* After the initial coding was discussed in detail, consensus coding was formed and the bridging approach was used to match transcribed citations to the analyzed coding, and quantitative findings were reinforced with qualitative findings. As shown in Tables 5a and 5b, all representative citations were mentioned by multiple people. Based on the qualitative analysis, we found support for our conclusions from the quantitative analysis, that flow experience and illusion of control will have some negative effects on users, which in turn will cause users to have negative emotions, which will ultimately affect users' intention to continue using short video application software.

Therefore, the results suggest that qualitative studies not only validate the structural choices of quantitative studies, but also draw similar conclusions to quantitative studies, thereby confirming the validity of quantitative inferences (Srivastava and Chandra, 2018; Venkatesh et al., 2016). The quantitative and qualitative data for this study came from different participants and different data collection processes, so the similarity of the findings suggests that this study has a strong theoretical structure (Srivastava and Chandra, 2018; Venkatesh et al., 2016). The richness and robustness of the results allow us to learn more about the usage survey of short video applications. Quantitative studies help us empirically test theoretical models, and qualitative studies further corroborate this.

*3.2.2.2. Qualitative inferences.* In addition to analyzing the same coding structure as the quantitative analysis through the interview data, we also have some new findings about users watching short videos.

Most of the respondents will watch short videos that are entertaining and funny. These short videos satisfy the respondents' desire for happiness and can relax their minds. There are also some respondents who watch short videos of games, food, current affairs news, film and television literature, cute pets, beauty, travel, fitness and shopping. Respondents pay attention to different types of short videos according to their own interests and learn about diversity in the world.

Among the 38 respondents, we found that 71 % of respondents believe that the frequency and duration of watching short videos has been affected by the outbreak of the COVID-19 pandemic. These respondents said that they have been watching short videos more frequently after the outbreak and watch them for longer. There are two main reasons, firstly some respondents work from home or study online most of the time, and secondly some are not allowed to travel due to restrictions. Among these respondents, some respondents started to watch short videos after the COVID-19 epidemic, which is consistent with the rapid development of the short video industry from 2020. 29 % of the respondents believe that the outbreak of the COVID-19 epidemic, has had no effect on their use of short video apps. Such respondents believe that their life and study are proceeding step by step, so they have little impact on themselves. Commenting on this issue, two interviewees remarked:

"I didn't watch short videos before, and Dou yin was downloaded after the outbreak of the COVID-19 epidemic. After the outbreak of the COVID-19 epidemic, I began to watch short videos." (A3).

*"After the outbreak of the COVID-19 epidemic, I will often watch short videos. One reason is that there is more time at home, and the other is that I think short videos have become more popular since 2020." (A13)*

When respondents were asked if they agreed with the idea that watching short videos is always fun, 80 % of the respondents replied that they did not agree with this sentence, they think this sentence is one-sided. Watching short videos can indeed bring happiness and excitement to people in a short period of time, but after a long period of time to a certain extent, users will feel that the happiness of watching short videos has decreased, and negative emotions such as fatigue, emptiness, regret, anxiety, etc. will appear. Respondents believe that the reason why users are easily addicted to short videos is that the short video recommendation mechanism exploits the weakness of human nature. Most of the respondents believe that in order to avoid getting themselves into anxiety and regret, it is still necessary to moderate watching short videos and not to be overly addicted, otherwise the extreme will be reversed. Commenting on this issue, as one interviewee remarked:

*"I think the experience of watching short videos is an inverted U shape. When I start watching, I feel happier and happier. At this point, I feel that watching short videos is good for expanding knowledge or decompressing. However, once I broke through the peak of excitement. The more I see, the more painful." (A1)*

Respondents expressed suggestions or opinions on the short video platform currently used, and most respondents said that the short video platform has inconvenient and unreasonable areas that need to be improved. Among them, the most important thing is to strengthen content review and supervision. Most of the respondents said that the platform should strengthen content supervision, should take the responsibility of correct content guidance, should resist pet abuse, vulgar broadcasters and deceived consumption, etc., should protect young people, protect Personal Intellectual Property Rights and creative achievements. Secondly, the repetition rate of short video push content is too high, which causes users to be troubled by the Information Cocoons (Meral, 2021), making it more and more boring for users to watch short videos. The third is that respondents believe that too many push advertisements, or too many soft advertisements, are embedded in short videos to affect the viewing experience. Some respondents also suggested that short video platforms should increase the anti-addiction mechanism for adults to prevent adults from indulging in short videos and causing addictive behavior. Finally, some respondents said that since the short video platform has such a large user base, from both the social level and the platform perspective, every-one should be encouraged to transmit positive energy through short video content, and promote Chinese culture. For example, two respondents remarked:

*"The short videos recommended by Dou yin are too repetitive and I feel boring after a while. So, the time I spend on Dou yin is getting shorter and shorter." (A11)*

*"I think short video platforms should strengthen content supervision and improve recommendation algorithms. Short video platforms should reduce the recommendation of negative video content, rather than blindly recommending things that users like in order to retain users. If users like vulgar content and explicit videos, there should also be positive guidance." (A19)*

*3.2.2.3. Meta-inferences: complementarity.* Next, we will integrate quantitative and qualitative findings through meta-inferences. We use the bracketing approach, which is the process of including respondents' differing opinions, to explore interesting phenomena and gain more complete insights. Meta-inferences with the expectation of further determining the boundary conditions of theoretical models for quantitative research (Cheng et al., 2022; Lewis and Grimes 1999; Venkatesh et al. 2013; Srivastava and Chandra, 2018). The following are three boundary conditions.

Boundary condition #1: Efficiency affects or not. Most of the respondents believe that short videos will affect their study/work efficiency. However, some respondents believe that short videos are a way to relax themselves. They think that as long as they can effectively control the time that they spend watching short videos and use the short video platform as a tool to relax after study/work, then it will not affect them. Efficiency, which in turn can improve learning/work efficiency. One such respondent remarked:

*"I will not completely reject short videos. I think if you can properly control the use of short video applications time is good for maintaining a good mood and releasing stress." (A20)*

Boundary condition #2: Time Awareness. Most of the respondents felt that the time passed quickly when they watched short videos and they regretted wasting time after watching them. Therefore, most respondents will consciously control the time they spend on short videos. However, some respondents do not set a time limit for themselves when watching short videos and they think they do not need to control it deliberately. This actually depends on how sensitive the user is to the passage of time and how concerned they are about time utilization. As one interviewee commented:

*"I didn't pay attention to the time when I was watching short videos. When I was bored, I watched short videos for a while." (A8)*

Boundary condition #3: Mindset Influence. Sadness is a kind of negative emotion experienced after users use short video application. Most of this negative emotion is generated in two situations, one is inflicted by negative short video content while they are watching, and the other is when users finish watching, and frustration and sadness is caused because they realize the delay that was caused to them doing their own work and other activities. However, some respondents said that this mainly depends on their mentality when watching short videos. Some respondents said that since wasting time has become a fact, they should accept it calmly and stop recalling what happened before. As one of the interviewees remarked:

*"Watching short videos is something I decide to do. If the short videos have any impact on my life, I think it should be borne by myself. Therefore, I can calmly accept the impact of short videos." (A35)*

## 4. Discussion

This study uses flow theory and illusion of control theory to develop a research model to link the negative factors brought about by users' short video watching with users' negative emotions and intention to continue using short video applications. Understanding the negative effects of short videos on users is conducive to the direction of the subsequent development of short video platforms and to meet the deep-seated needs of users.

This study adopts a mixed research method. In the first stage the research model was developed. It showed how the (1) flow experience and (2) illusion of control, influenced (3) negative emotions, which finally influenced (4) the continuous usage intention of short videos. The analysis of 420 surveys supports the model. In the second stage, this paper conducted 38 semi-structured interviews and verified the quantitative analysis inferences, as well as supplemented the quantitative analysis results. At the same time, it enriched this research with users' insights into short videos. This research tested the six antecedent factors identified by the model: low efficiency, time distortion, harm to health, online addiction, online procrastination and illusion of control. Through quantitative analysis, it is concluded that the flow experience brought by the user's immersive short video watching, and the illusion of control that happens when the user believes they control their use of short video, have a significant impact on the user's emotions. More specifically, the negative effects of low efficiency, time distortion and online procrastination brought to users by short videos will lead to negative emotions of remorse for users. The negative effects of harmful physical and mental health, online addiction, online procrastination and the illusion of control can lead to negative feelings of anxiety in users. In addition, the negative effects of harmful physical and mental health, online addiction and illusion of control can lead to sad negative emotions in users. In other words, the damage to users' physical and mental health caused by using short video platforms can lead to feelings of anxiety and sadness, as well as online addiction and the illusion of control. Online procrastination can create feelings of regret and anxiety in users. Furthermore, low efficiency and time distortion can also cause user to feel remorse.

In addition, quantitative analysis also found that users' negative emotions caused by watching short videos have a conspicuous affected on users' intention to continue to use short video platforms. Among them, remorse is negatively related to users' intention of continuing to use short video platforms and sadness is positively related. Moreover, anxiety has no effect on users' intention to continue using short video platforms. This finding is reasonable. Most users will regret wasting their precious time after watching short videos, so they think about controlling themselves not to watch short videos. However, the user's sad mood lasts for a shorter period than the regretful mood, or the user wants to temporarily relieve the depressed mood and obtain short-term happiness by watching short videos. Therefore, remorse makes users reduce their intention to continue using the short video platform, but sadness makes users increase their intention to continue using the short video platform. Users will have anxiety when using short video platforms, but anxiety will not affect users' intention to continue using short video platforms. That is to say, some users will reduce their intention to continue using short video platforms because of anxiety, while others will rely more on short video platforms because of anxiety. The inference that negative emotions affect users' intention to continue to use short video platforms is verified in the qualitative interview data. Firstly, the respondents who have negative emotions due to the use of short video platforms indicate that the time to watch short videos will gradually decrease in the future. Secondly, such respondents also expressed their reluctance to download more short video applications. The existing applications or platforms are enough to watch short video content, because short video forms have been flooded in major application software. In the end, such respondents said they were reluctant to recommend short video platforms to their relatives and friends, because apart from the fact that short videos are too popular and do not need to be recommended, respondents believe that short videos do more harm than good and do not want to let the people around them suffer from it.

Furthermore, through quantitative research, it is found that the control variables are also remarkable on the intention of users to continue to use the short video platforms. Age and daily watching time of short videos are positively related to the intention of continuing to use short video platforms, and education level is negatively related to the intention of continuing to use the platform. This finding shows that with the increase of age, users are more willing to continue to use short video platforms, which means that users should be discouraged from over-using short videos, and they should be encouraged to spend their time on more valuable activities. However, older users have a more stable work and social life, and some of their spare time can be used on short videos with less risk of overusing them. Even if there are some negative effects, it cannot affect their intention to continue to use short video platforms. The more users watch short videos every day, the more dependent they are on short videos, and the stronger their intention to continue using short video platforms. The more time users spend on short videos every day also means the proportion of their day spent on short videos increases. The more educated users have more self-control and stricter time management, they don't want to waste time on short videos. However, the less educated users have more ambiguous concepts of time or have more entertainment time at their disposal. The conclusions of quantitative analysis of control variables were also verified in qualitative analysis. By analyzing the age, education level and how much time is spent on watching short videos every day, the 38 respondents of the qualitative section had the same beliefs as the quantitative analysis.

Finally, in the extended question of the qualitative analysis, we learned whether the frequency and duration of users' short videos have been affected by the COVID-19 epidemic in recent years. 71 % of the respondents said that they were affected, while 29 % of users said that they were not affected. In addition, through qualitative analysis, it is also found that users do not feel that watching short videos is always a happy experience, and most users can maintain rationality and self-discipline. Finally, I also learned users' suggestions on the short video platform they are currently using. In addition, this paper conducts boundary condition research from three aspects: whether the user's perception of short video affects their efficiency, the user's perception of time, and how the user's mentality affects the user's emotion. From these boundary conditions, it can be seen that each person's experience of using the short video platform may be different at each stage, which is related to the user's current environment and psychological state.

In conclusion, this paper explores the negative effects and negative emotions of users caused by using short video platforms, and analyzes the relationship between negative effects and negative emotions and users' intentions to continue to use short video platforms through research models using quantitative research and qualitative research. Studies have shown that users' use of short video platforms has six negative effects (low efficiency, time distortion, harm to health, online addiction, online procrastination, illusion of control) and three negative emotions (remorse, anxiety, sadness). And the research conclusion shows that negative emotions affect users' intention to continue to use short video platforms. Specifically, the health damage, online addiction and illusion of control caused by users' use of short video platforms can cause users to feel anxious and sad. Online procrastination can cause remorse and anxiety, and low efficiency and time distortion can cause users to feel remorse. User's remorse will reduce the user's intention and sadness will enhance the user's intention to continue to use the short video platform. However, anxiety did not have a significant impact on users' intention to continue to use, indicating that users' anxiety did not affect their intention to continue to use the short video platforms. In addition, the user's age and the length of daily use of the short video platform have a positive impact on the user's intention to continue to use the short video platform, and the user's education level has a negative impact on the user's intention to continue to use the short video platform.

## 5. Conclusion

### 5.1. Theoretical implications

The results of this study have three theoretical contributions. First, the research in this paper focuses on the negative effects and negative emotions of users using short video platforms, and users' intention to continue using short video platforms when they have negative emotions. This direction was chosen because of few scholars have studied deeply the negative emotions generated by the negative effects of short video platforms on users (Gao et al., 2017; Hong et al., 2014; Huang et al., 2022; Ye et al., 2022). Furthermore, in the context of the COVID-19 epidemic, we learned through qualitative research that the duration and frequency of users' use of short video platforms are affected by the COVID-19 epidemic. Therefore, this study expands the breadth and depth of research on short videos and enriches the study of negative emotions on the intention to continue using human–computer interaction software.

Second, this study uses flow theory (Csikszentmihalyi, 1975a) and illusion of control theory (Langer, 1975) to create a research model combined with negative emotions. This model deeply explores the relationship between the negative impact (Huang et al., 2022; Ye et al., 2022), negative emotions and continuous usage intentions of users using short video platforms. This study expands the research on the negative effects of flow theory, and at the same time expands the research on flow experience in the field of short video, making the application scenarios of flow theory more extensive. In addition, existing literature mainly applies illusion of control theory to experimental and gambling research (Bandera et al., 2018; Fu and Yu, 2015; Meng and Leary, 2021; Yu and Fu, 2019). In this study, the illusion of control theory is introduced into the field of human–computer interaction for the first time, which enriches the application scenarios of control illusion theory and fills the gap of control illusion theory in the field of human–computer interaction.

Finally, this study uses a mixed methods research approach (Venkatesh et al. 2013; Venkatesh et al. 2016) and shows how an explanatory sequential design (a combination of quantitative and qualitative research) can be used to draw rich and meaningful conclusions and meta-inferences (Srivastava and Chandra, 2018). Specifically, we analyze quantitative research results and confirm the inference correctness of quantitative research with the bridging approach (Lewis and Grimes, 1999; Venkatesh et al., 2013). Furthermore, we supplemented the findings by analyzing the interview data to enrich the content not covered by the quantitative analysis. Next, we supplement the quantitative analysis results by using the bracketing approach (Lewis and Grimes, 1999; Venkatesh et al. 2013; Srivastava and Chandra, 2018) and derive three boundary conditions that give us a deeper understanding of the conditions that limit quantitative conclusions through qualitative analysis. The identified boundary conditions contribute to the overall understanding of the conclusions drawn.

### 5.2. Practical implications

As information becomes more open and transparent, the trend of using short video platforms as information transmission carriers will intensify. This research confirms some negative effects and negative emotions brought by short videos to users, which in turn affect users' intention to continue using short video platforms. Therefore, we provide some practical suggestions for managers of short video platforms, and these same suggestions also provide some references for policy makers of short video platforms.

First, managers should weigh the pros and cons of short video platform recommendation algorithms for users. Short video platforms use recommendation algorithms to firmly grasp users' watching preferences, but they do not know that extremes can be counterproductive. In the long run, this can lead to negative effects and negative emotions for users. Therefore, short video platforms should pay more attention to the user's resistance to the recommendation algorithm (Velkova and Kaun, 2021). Managers should optimize the recommendation algorithm to find a recommendation degree suitable for users and enrich the diversity of recommendations.

Second, managers should strengthen the anti-addiction mechanism of short videos, not only for teenagers, but also for adults. This also applies in the COVID-19 epidemic that influenced daily life and people have more time for short video platform surfing. Now, users not only hope to get happiness and information in short videos, but also hope that the short video platform will not affect their normal study and work. At present, many users feel remorse, anxiety and sadness because they are immersed in short video platforms. Short video technologies strive to increase the usage time of users, but if this goes too far the user will stop using them. Therefore, short video platforms should avoid the negative effects and negative emotions of users due to the use of short video platforms, so that users can continue to use them.

Finally, managers should fully respect human nature and not take advantage of human nature. Short video platforms should not aim at eroding users' time and money, but should also respect the characteristics of users' humanity and give users a blue ocean of rationality. This study also offers implications to the policy makers that can develop new policies so that the short video platform economy can develop in a positive way for all the stakeholders.

### 5.3. Limitations and future research

This study still has some limitations. The data from the interviews are self-reported by respondents, so data can be biased by social expectations. However, we did not control for potential respondent response bias by designing and administering questionnaires and interview questions (Reisman et al., 2006). In the future, we can conduct individual surveys of respondents through survey voluntariness and management systems that control social expectations bias (Nederhof, 1985). Moreover, our research model uses cross-sectional data. In other words, users' perceptions of watching short videos and their intention to continue using short video platforms were measured at a single point in time. The results of the study may be affected by the respondents' recent mentality and life status. We also need further longitudinal studies to complement and extend our findings. We also may need to test our findings in other settings, such as more AI enabled communication and collaboration platforms.

**Declaration of Competing Interest**

The authors declare that they have no known competing financial interests or personal relationships that could have appeared to influence the work reported in this paper.

**Data availability**

The authors do not have permission to share data.

**Acknowledgements**

We would like to thank National Natural Science Foundation of China (Grant No. 72061147005 and 72271236), the School of Interdisciplinary Studies in Renmin University of China, Metaverse Research Center in Renmin University of China for providing funding for part of this research.

# Appendix A. Measurement scales

| Construct | Item | Content | Resource |
|---|---|---|---|
| Low efficiency | LE1 | After indulging in watching short videos, I feel that my work/study efficiency is reduced. | Lam et al. (2009) |
| | LE2 | After indulging in watching short videos, I feel that my patience to complete complex tasks has dropped. | |
| | LE3 | After indulging in watching short videos, I feel less interested in my work/study. | |
| | LE4 | After indulging in watching short videos, I feel that the number of tasks done in work/study has decreased. | |
| Time distortion | TD1 | Time seems to pass quickly when I'm immersed short videos. | Koufaris (2002); Novak et al. (2000) |
| | TD2 | I tend to lose track of time when I'm immersed short videos. | |
| | TD3 | I frequently watch short videos longer than originally intended. | |
| Harm to health | HH1 | After "immersive" watching short videos, I feel that I can't calm down for a long time. | Zhang et al., 1993 |
| | HH2 | I feel lost after watching short videos in "immersive". | |
| | HH3 | I feel vast and hazy after watching short videos in "immersive". | |
| | HH4 | I feel empty after watching short videos in "immersive". | |
| Online addiction | OA1 | I have tried several times to control not watching short videos, but they all ended in failure. | Young (1998) |
| | OA2 | When I try to control myself from watching short videos, I feel restless, moody, depressed, or irritable. | |
| | OA3 | I have lost important friends, jobs or education opportunities because I was addicted to watching short videos. | |
| Online procrastination | OP1 | I'm immersed in watching short videos to avoid work or study tasks. | Thatcher et al. (2008); Hernández et al. (2019) |
| | OP2 | Frequently, I'm immersed in watching short videos to postpone tasks that I find unpleasant or difficult. | |
| | OP3 | I tend to procrastinate when watching short videos. | |
| Illusion of control | IC1 | I believe I can control myself from watching short videos. | Moore and Ohtsuka (1999); Fu and Yu (2013) |
| | IC2 | I believe I can control how long I watch short videos every day. | |
| Remorse | R1 | I would blame myself for watching short videos for too long. | Watson et al. (1988) |
| | R2 | I will feel guilty after watching short videos. | |
| | R3 | I will feel remorse after watching short videos. | |
| | R4 | I will blame myself for spending too long on short videos. | |
| Anxiety | A1 | I will feel uneasy after watching short videos. | Zung (1965) |
| | A2 | I will feel irritable after watching short videos. | |
| | A3 | I will feel tired at heart after watching short videos. | |
| Sadness | S1 | I will feel unhappy after watching short videos. | Watson et al. (1988) |
| | S2 | I feel low for a while after watching short videos | |
| | S3 | I will feel depressed after watching short videos. | |
| Continuous usage intention | CUI1 | I'll download more short video apps in the future. | Bhattacherjee (2001) |
| | CUI2 | I will spend more time watching short videos in the future. | |
| | CUI3 | I would like to recommend the short video application software to my relatives and friends. | |